\documentclass[aps, prb, onecolumn, superscriptaddress, amsmath, amssymb]{revtex4}

\usepackage{graphicx}
\usepackage{dcolumn}
\usepackage{bm}
\usepackage{caption}
\usepackage{layouts}
\usepackage{xcolor}

\begin{document}

\title{Atomic-scale modeling of superalloys
\footnote{Chapter 11 of \emph{Nickel Base Single Crystals Across Length Scales}, Eds. G. Cailletaud et al., Elsevier 2022}}

\author{Thomas Hammerschmidt}
\affiliation{ICAMS, Ruhr University Bochum, D-44801 Bochum, Germany}
\author{Jutta Rogal}
\affiliation{Department of Chemistry, New York University, New York, NY 10003, USA and Fachbereich Physik, Freie Universit{\"a}t Berlin, D-14195 Berlin, Germany}
\author{Erik Bitzek}
\affiliation{Department of Materials Science and Engineering, Institute I, Friedrich-Alexander-Universit{\"a}t Erlangen-N{\"u}rnberg, D-91058 Erlangen, Germany}
\author{Ralf Drautz}
\affiliation{ICAMS, Ruhr University Bochum, D-44801 Bochum, Germany}

\date{\today}

\begin{abstract}
Atomistic theory holds the promise for the ab initio development of superalloys based on the fundamental principles of quantum mechanics. The last years showed a rapid progress in the field. Results from atomistic modeling enter larger-scale simulations of alloy performance and often may be compared directly to experimental characterization.
In this chapter we give an overview of atomistic modeling and simulation for Ni-base superalloys. 
We cover descriptions of the interatomic interaction from quantum-mechanical simulations with a small number of atoms to multi-million-atom simulations with classical interatomic potentials. Methods to determine structural stability for different chemical compositions, thermodynamic and kinetic properties of typical defects in superalloys, and relations to mechanical deformation are discussed. Connections to other modeling techniques are outlined.
\end{abstract}

\maketitle

\section{Introduction}\label{sec:atomistic-intro}

In 1929 Paul Dirac noted that ``The underlying physical laws necessary for the mathematical theory of a large part of physics and the whole of chemistry are thus completely known, and the difficulty is only that the exact application of these laws leads to equations much too complicated to be soluble'' \cite{Dirac29}. He continued to say that ``It therefore becomes desirable that approximate practical methods of applying quantum mechanics should be developed, which can lead to an explanation of the main features of complex atomic systems without too much computation.''

Three aspects of Dirac's statement largely determine our work today. The first observation is that the fundamental laws, for the modeling of materials with the many-electron Schr\"{o}dinger equation, are known. As ultimately the behavior of a material is controlled by the bonds between atoms that are mediated by electrons, we are therefore in principle in a position to understand, predict, and design materials from first principles. 
The caveat, and Dirac was very clear about this in his statement, is that it is strictly impossible to solve the many-electron Schr\"{o}dinger equation for a superalloy. Therefore today, nearly a century after Dirac's statement, we are still developing approximate models that are rooted in the fundamental laws of nature to support materials design from first principles.

A breakthrough was made by Kohn and coworkers when they developed density functional theory (DFT) \cite{Hohenberg64,Kohn65}, today easily the most cited concept in the physical sciences \cite{top100}. The effective one-electron structure of DFT makes it amenable to further approximations that treat the electrons as a mere glue between the atoms, so that effective classical interatomic potentials may be derived.

\begin{figure}[h]
\includegraphics[width=0.9\columnwidth]{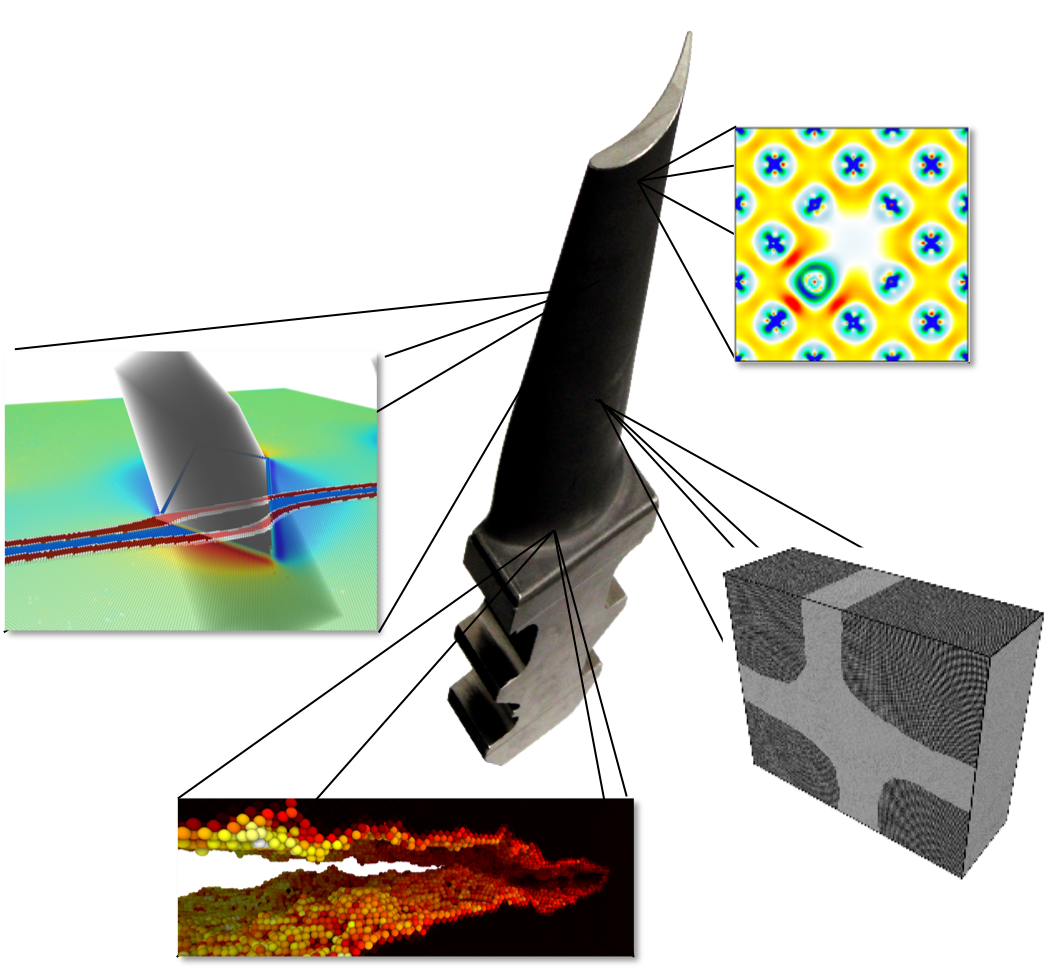}
\caption{Aspects of superalloy turbine blades relevant to development, use, and failure that can be addressed by atomistic simulations. From top to bottom: vacancy formation energies as functions of local chemical composition; interaction of a superdislocation in $\gamma'$ with a $\gamma$ precipitate and resolved misfit stresses; simulation of $\gamma $/$\gamma'$-microstructures; fracture.}
\label{fig1:intro}
\end{figure}

In this chapter, we summarize the main aspects of the present state-of-the-art of atomistic modeling and simulation for superalloys. We start from small calculations with only a few atoms, for which DFT may be employed, and end with atomistic simulations of the microstructure that require many millions of atoms and that are carried out with classical interatomic potentials, see Fig.~\ref{fig1:intro}. We list limitations that need to be overcome for the ab initio development of superalloys.

\section{Methods}\label{sec:atomistic-methods}

\subsection{Modeling atomic interactions}

The structural stability and mechanical properties of Ni-base superalloys are driven by the distribution of the different chemical elements and the features of the microstructure. 
Atomistic modeling therefore needs to account for the interaction of the complex alloy chemistry and the diverse geometric features of the atomic structure of imperfections like point defects, interfaces, dislocations, stacking faults, and precipitates. 
This requires capturing the atomic interactions with appropriate accuracy and tackling simulation cells with a sufficiently large number of atoms. 
These are antipodal requirements as higher accuracy requires higher complexity and computational effort and therefore leads to smaller simulation cells. 
The hierarchy of approaches discussed in the following spans the range from highly precise quantum-mechanical calculations at nm length-scales to classical simulations which nowadays can attain $\mu$m length-scales.

The most accurate approach discussed here involves DFT calculations that provide a numerical solution to the quantum-mechanical equations for the interaction of electrons and ions. 
The results for metallic systems like Ni-base compounds are highly reliable but the computational effort sets a practical limit to simulation cells with typically a few hundred atoms. 
This quantum-mechanical description can be simplified in a second-order expansion of DFT \cite{Drautz-11} to the tight-binding (TB) bond model \cite{Sutton-88}. 
This increases the tractable system size to several thousand atoms at the expense of reduced accuracy and an additional parameterization step. 
The required reference data for the latter are typically energies and forces of crystal structures and defects obtained by DFT calculations. 
The numerical solution of the simplified quantum-mechanical description can be further accelerated by bond-order potentials (BOPs) (see, e.g., \cite{Drautz-15}). 
The underlying approximation effectively localizes the range of the quantum-mechanical interaction to a few neighbor shells.
With the resulting linear scaling of computational effort with the number of atoms and the possibility for efficient parallelization, the tractable number of atoms is extended to several hundred thousands in a routine simulation \cite{Hammerschmidt-19}. 
A further level of simplification and millions or even billions of atoms can be reached by giving up even the simplified quantum-mechanical description of TB/BOP for a set of short-ranged, analytical functions that mimic the interatomic interaction. 
This allows constructing approximate models with simple mathematical forms that are parameterized for specific chemical elements and atomic structures. 
For metallic systems like Ni-base superalloys, pair potentials like the Finnis--Sinclair potential \cite{Finnis-84} and embedded-atom method \cite{Daw-93} are particularly successful and have been parameterized for studying $\gamma $\;and $\gamma'$\;phases of NiAl~\cite{Mishin-04,PurjaPun-09}. 
A more recent development are machine-learning potentials for superalloys \cite{Chandran-18} that replace the fixed set of short-ranged mathematical functions with artificial intelligence approaches in order to interpolate large sets of DFT data for the atomistic description of the superalloy.

\subsection{Calculation of structures and energies}

The atomistic methods given above start by constructing an atomistic representation of the particular aspect of the superalloy that should be studied. Using the positions and chemical species of the atoms, the different atomistic simulations methods (DFT, TB, BOP, classical potentials) deliver the potential energy of the system, the forces on the individual atoms, and the stresses on the simulation cell.

One of the most basic examples is the computation of the lattice constant of an Ni$_3$Al $\gamma'$\;phase.
The corresponding atomistic representation is a unit cell of an fcc (face-centered cubic) crystal lattice with an L1$_2$-ordered occupation by Ni and Al atoms in a 3:1 ratio subject to periodic boundary conditions.
The computation of the total energy with one of the atomistic methods above depends on the input lattice constant that is chosen for the unit cell.
The lattice constant where the total energy takes its minimum value is the equilibrium lattice constant and the corresponding energy is the equilibrium energy.
The application of strain tensors to the equilibrium unit cell leads to stresses which can be used to determine the elastic constants of the material.

The defect structures discussed in the following sections are represented atomistically by repetitions of the unit cell and additional modifications.
For example, a $\gamma'$\;phase of Ni$_3$Al with 0.2\,at.\% Ni$_{\textrm{Al}}$ antisite defects is obtained by a five-fold repetition of the L1$_2$-Ni$_3$Al unit cell with four atoms in [100], [010], and [001] directions followed a replacement of one Al atom by an Ni atom.
Similarly, the site preference of alloying elements (see, e.g., \cite{Jiang-06}) and their influence on structural stability and elastic constants (see, e.g., \cite{Kim-10}) can be determined by corresponding total-energy calculations where part of the Ni or Al atoms are replaced by the alloying elements of interest.

In these cases, the simulation cell has more degrees of freedom than only the lattice constant, and a structure relaxation needs to be performed.
This minimization of the total energy with respect to all structural degrees of freedom leads to a refinement of unit cell size and shape, and atomic positions.
The central results of the relaxation of such structures are the detailed atomic structure including displacements and the energetics in the limit of $T=0K$.
This is the starting point for further simulations of finite-temperature effects and defect mobility discussed in the next sections.
Examples of large-scale simulation setups for $\gamma $/$\gamma'$ structures, precipitate cutting and fracture are given in Fig.~\ref{fig1:intro}.

\subsection{Finite temperature calculations}\label{sec:finiteT}

Total energy calculations as discussed in the previous section provide valuable insight into the structural properties and relative stability of metal alloys. The effect of temperature is, however, not included in these calculations. To be able to compare thermodynamic properties at finite temperatures, it is necessary to determine the different entropy contributions to the free energy. Furthermore, additional simulation tools are required to study the dynamical properties on an atomistic level.

Assuming an adiabatic decoupling of the different degrees of freedom, the free energy can be expressed as
\begin{align}
\label{eq:freeenergy}
F(V,T) = E^{\text{tot}}(V) +\! F^{\text{el}}(V,T) +\! F^{\text{vib}}(V,T) +\! F^{\text{mag}}(V,T) +\! F^{\text{conf}}(V,T) + \cdots\! ,
\end{align}
where $E^{\text{tot}}$ is the total energy at $T=0K$, $F^{\text{el}}$ is the electronic, $F^{\text{vib}}$ the vibrational, $F^{\text{mag}}$ the magnetic, and $F^{\text{conf}}$ the configurational contribution to the free energy. The coupling between different degrees of freedom (electronic--vibrational, vibrational--magnetic) would give rise to extra terms in this expansion. Furthermore, any type of defect (like vacancies) would additionally contribute to the free energy.
The electronic free energy, $F^{\text{el}}$, for bulk systems can be computed with high accuracy based on ab initio calculations \cite{Zhang17}, where methods assuming fixed atomic positions include the self-consistent finite temperature DFT approach, the fixed density of states approximation, and the Sommerfeld approximation.
The free energy of atomic vibrations, $F^{\text{vib}}$, usually constitutes the largest contribution at finite temperatures \cite{Palumbo14}. Within the harmonic approximation, the vibrational free energy is calculated from the phonon frequencies which are accessible within a first-principles approach either by density functional perturbation theory \cite{Baroni01} or by the small displacement approach \cite{Kresse95}.
To first order, anharmonicity can be included via the quasiharmonic approximation \cite{Palumbo14}. Here, the dependence of the phonon frequencies on the volume is explicitly considered. At each temperature, the free energy is then accessible as a function of volume, where the minimum denotes the equilibrium volume at the corresponding temperature.
Calculating vibrational free energies including full anharmonicity can be achieved on the basis of molecular dynamics (MD) simulations, discussed below. A widely used approach is thermodynamic integration \cite{FrenkelBook} between a system with known free energy (e.g., within the quasiharmonic approach) and the system of interest. MD simulations based on energies and forces from ab initio calculations (AIMD) can provide very accurate results, but are also computationally very demanding.
Including magnetic contributions to the free energy, $F^{\text{mag}}$, from first-principles calculations is rather challenging \cite{Koermann16}, usually requiring the setup of model Hamiltonians derived from ab initio calculations. These contributions are, however, important in magnetic transition metals such as Ni, Co, and Fe, where they significantly impact thermodynamic quantities such as the free energy and the specific heat \cite{Koermann11}.
In disordered alloys, the last term, $F^{\text{conf}}$, is often approximated using the entropy of mixing for an ideal solid solution. More sophisticated approaches have been developed based on a cluster expansion approach that aim to sample the space of all possible arrangements of different atom types on a fixed crystal lattice \cite{VandeWalle07,Zhang16}.
Together with Monte Carlo (MC) sampling approaches in different thermodynamic ensembles it becomes possible to investigate phenomena such as order-disorder transitions, precipitate formation, or phase diagrams \cite{Goiri16,Maisel14}.

To investigate the dynamical behavior of materials at finite temperatures, MD simulations \cite{FrenkelBook,AllenBook} constitute nowadays the standard workhorse for atomistic simulations. In MD, atoms are treated as classical particles following Newton's equations of motion. Simulations can be performed in various thermodynamic ensembles by using the corresponding thermostats and barostats, including the microcanonical, canonical, and isothermal--isobaric ensembles. Advances in numerical algorithms and increasing computational resources allow performing simulations with up to billions of atoms. The development of elaborate program packages \cite{lammps} has contributed to a widespread use of MD simulations to study materials properties.
The two main difficulties in MD simulations are an accurate description of the interatomic interactions and the accessible timescales. Ab initio MD \cite{MarxBook} is based on highly accurate energies and forces, but system sizes are limited to a few hundred atoms. Empirical potentials, on the other hand, are usually not reliable for complex, multicomponent alloys and often have difficulties to properly represent complex structural environments, such as dislocation cores or during crack propagation. The timescale problem in MD simulations arises when the process of interest involves sizeable energy barriers between two metastable states of the system, e.g., during vacancy mediated diffusion (see also Section~\ref{sec:atomistic-pointmobility}).
This is due to a separation of timescales between the fast vibrations and the comparably slow changes in structure or exchange of atomic positions. A number of accelerated MD techniques \cite{Voter02} has been developed to facilitate the escape from metastable states in the course of the simulation while preserving the correct dynamics. Another approach is kinetic Monte Carlo (KMC) \cite{Bortz75,Fichthorn91} simulations where a rate constant is determined for each process connecting two metastable states and the dynamics is given by a stochastic state-to-state trajectory that again preserves the correct timescale. If a suitable mapping of the investigated system onto a finite state space can be identified, then the corresponding rate constants for all processes can be determined with high accuracy based on electronic structure calculations.
The analysis of dynamical simulations provides insight into the atomistic mechanisms and gives access to macroscopic transport coefficients via the corresponding Green--Kubo relations \cite{FrenkelBook,AllenBook}, as, e.g., diffusion coefficients, shear viscosity, or thermal conductivity. These quantities can directly be compared to experimental measurements.

\section{Thermodynamic stability}

A central concept for determining the structural stability of crystalline phases by atomistic simulations are differences of total energies from simulations for different systems.
As an example, the formation energy of the Ni$_3$Al $\gamma'$\;phase with respect to the elemental ground states is taken as the energy difference between the L1$_2$ Ni$_3$Al total energy and the total energy of fcc-Al and fcc-Ni where each total energy is obtained in a separate simulation. With proper accounting of the chemical compositions, one can compare different crystal structures with different chemical composition by convex-hull constructions. The resulting information on the energetically most favorable crystal structure for a given chemical composition at $T=0K$ and further entropy contributions can be considered in CALPHAD assessments of phase diagrams of compound systems, see Chapter~2.

The concept of energy differences is also used to investigate the stability of defects.
The segregation energy of an alloying element to a defect, e.g., is taken as energy difference between the total energy of a supercell with the alloying element positioned at the defect and a second supercell with the alloying element far away from the defect.
Similarly, the formation energies of point defects (e.g., vacancies), line defects (e.g., dislocations), planar defects (e.g., stacking faults), and extended defects (e.g., precipitates) are computed from the difference in energy of calculations with and without the corresponding defect.

A certain limitation of the atomistic approaches for the case of typical multicomponent superalloys is a realistic representation of their chemical complexity. Classical interaction models, on the one hand, usually lack reliable multicomponent parameterizations, and DFT calculations, on the other hand, are hardly possible due to (i) the size of the required supercells and (ii) the combinatorially large number of atom distributions within the supercell.
An alternative approach are structure maps \cite{Pettifor-84-2,Pettifor-86-1,Hammerschmidt-08,Seiser-11-1,Bialon-16} that chart the trends of the structural stability \cite{Pettifor-86-2,Seiser-11-2,Hammerschmidt-13} in low-dimensional representations.
The application of structure maps to superalloys could, e.g., rationalize the formation of detrimental topologically close-packed (TCP) phases in the CMSX4-like Ni-base superalloy ERBO/1 in the as-cast and heat-treated state \cite{Lopez-16,Hammerschmidt-16}, in low-cycle fatigue-tests \cite{Meid-19}, and during repair by vacuum plasma spray \cite{Kalfhaus-19}.

\section{Point defects}

In general, there are three types of point defect that we consider in bulk phases: vacancies, where an atom is missing from a particular lattice site; substitutional defects, where a matrix atom is replaced by another element; and interstitial defects, where additional atoms are present in between regular lattice sites.
In the following we will focus on the former two, vacancies and substitutional defects.

\subsection{Thermodynamic properties}

\begin{figure}[htb]
\includegraphics[width=0.9\columnwidth]{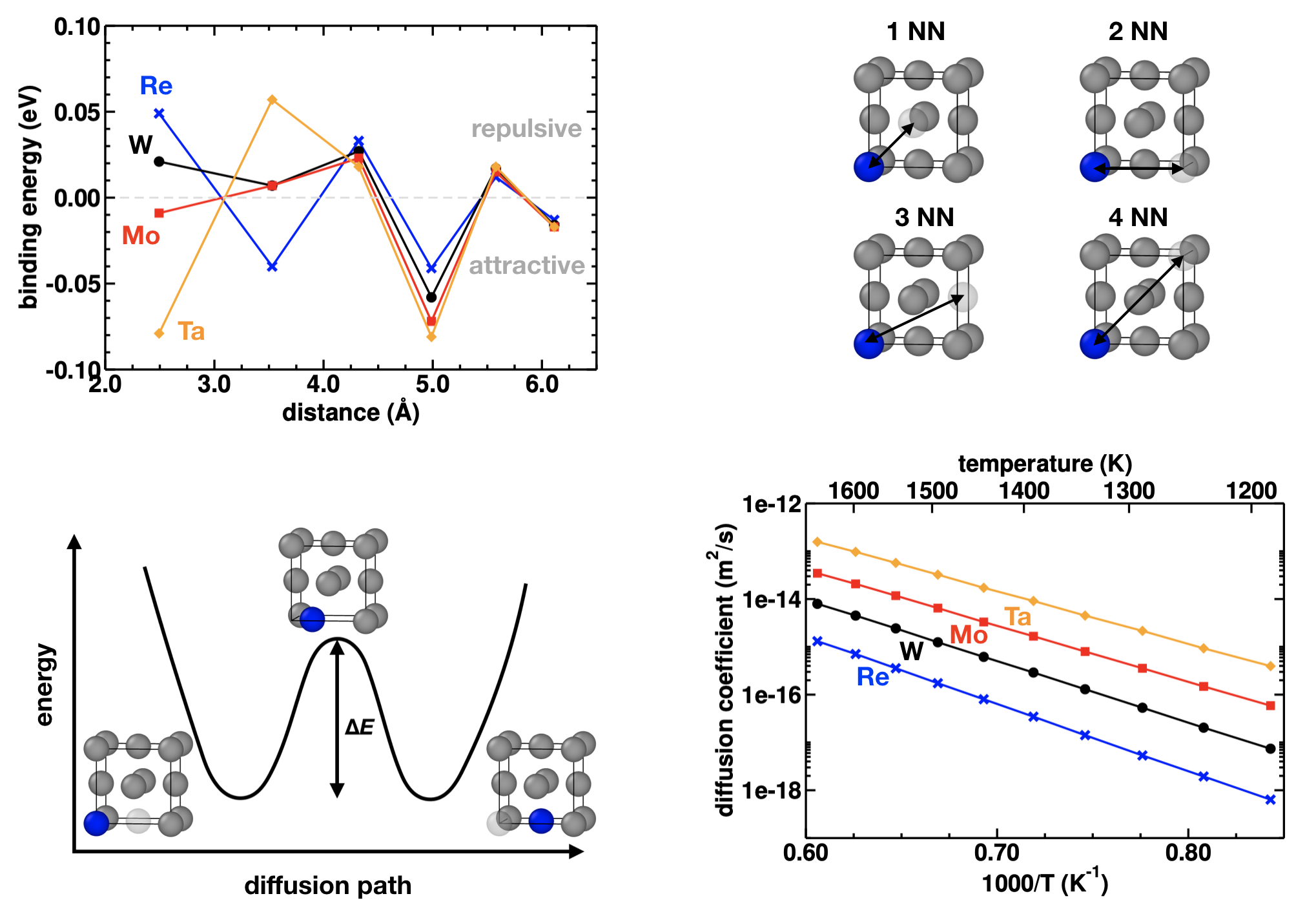}
\caption{(Top) Interaction energies between a vacancy and solute atoms (Re, W, Mo, Ta) in an fcc Ni matrix. Overall the interaction energies are small on the order of 50-100~meV. In the first (1NN) and second (2NN) neighbor shell, the interaction energies are dominated by electronic effects, whereas for larger distances elastic effects due to the difference in size between the solute atoms and Ni become important. In the first neighbor shell, the interaction between a vacancy and Re or W are repulsive, whereas the interaction with Mo or Ta are attractive. The observed trend in the interaction energies can be correlated with the filling of the $d$-band in the electronic structure of the solute atoms. Figure adapted from \cite{Schuwalow14}. 
(Bottom left) Schematic representation of the energy along the diffusion path of an atom (blue) exchanging its position with a vacancy (transparent); the energy difference between the initial and transition state corresponds to the energy barrier $\Delta E$ of the diffusion process. 
(Bottom right) Diffusion coefficients of alloying elements (Re, W, Mo, Ta) in fcc Ni as a function of temperature. The diffusion coefficients have been calculated using KMC simulations with barriers from DFT calculations. The diffusion-activation energy $Q$ can be extracted from the slope of the corresponding linear fit to the numerical data in the Arrhenius plot. Figure adapted from \cite{Schuwalow14}.}
\label{fig:Epoint}
\end{figure}

One of the important thermodynamic quantities is the defect formation energy, that is, the change in energy due to the creation of a defect. Atomistically, the defect formation energy can be computed within a supercell approach as the difference between the total energy of supercells with and without the corresponding defect, see \cite{Freysoldt-14} for a review. The vacancy formation energy (VFE) in pure Ni is, e.g., given by
\begin{equation}
\Delta E^{\text{VFE}} = E_{\text{Ni}_{X-1}\text{Va}} - \frac{X-1}{X} E_{\text{Ni}_{X}} ,
\end{equation}
where $X$ is the number of atoms in the supercell. The size of the supercell determines the minimum defect concentration and needs to be carefully tested to avoid artificial defect--defect interactions \cite{Rogal14}.
Similarly, the magnitude of the interaction between point defects can be calculated as the energy difference between an isolated defect, that is a single defect in a large supercell, and two or multiple defects at a certain distance. Attractive interactions indicate binding or clustering, whereas repulsive interactions favor a random distribution of point defects in the matrix.
In Fig.~\ref{fig:Epoint} the interaction energies between a vacancy and solute atoms (Re, W, Mo, Ta) in an Ni matrix are shown.
These calculations demonstrate that single Re atoms do not bind vacancies \cite{Schuwalow14,Zhang14a} and that this could be dismissed as a possible hypothesis for the Re effect.
Similarly, nonmagnetic electronic structure calculations of interaction energies between Re atoms in Ni revealed that Re does not tend to form clusters \cite{Mottura12}, eliminating another speculation concerning the cause of the Re effect. For the Ni--Re system the results do, however, strongly depend on the magnetic state of the system \cite{He16}.
In multicomponent systems, the defect formation energy depends on the local chemical environment as well as on the global composition. In random solid solutions, it is in addition unknown which atom type previously occupied the defect site, which requires considering a properly weighted average over possible configurations \cite{Rogal14}. In chemically ordered phases, the formation energy of antisite defects (an atom of one sublattice occupies a site on another sublattice) determines how the system compensates off-stoichiometric compositions. In the L1$_2$ ordered Ni$_3$Al $\gamma'$\;phase, the formation energy of antisite defects is much lower than the vacancy formation energy \cite{Koning02,Jiang06b,Gopal12,Zhang14b,Goiri16} and thus structural vacancies are not observed for off-stoichiometric compositions \cite{BaduraGergen97}. Furthermore, a number of first-principle studies have investigated the site preference of ternary alloying elements in the $\gamma'$\;phase \cite{Jiang06b,Ruban14,Ruban97,Kim12},
which was shown to influence the mechanical properties of this phase \cite{Rawlings75}.

\subsection{Mobility}

\label{sec:atomistic-pointmobility}
Solid state self-diffusion and the diffusion of substitutional defects are mainly mediated by vacancies. On the atomic scale, a diffusion process involves the exchange of an atom with a neighboring vacancy which is associated with an energy barrier as shown in Fig.~\ref{fig:Epoint}.
The minimum energy path along the diffusion process can be determined using the nudged-elastic band (NEB) approach \cite{Henkelman00} or the string method \cite{E07} with a high degree of accuracy using electronic structure methods. The corresponding microscopic diffusion barriers can either be used in analytical models \cite{LeClaire56,Goswami14} or as input parameters to kinetic Monte Carlo (KMC) \cite{Bortz75,Schuwalow14} simulations to calculate macroscopic diffusion coefficients. In Fig.~\ref{fig:Epoint} diffusion coefficients of alloying elements in Ni as a function of temperature are shown determined by DFT calculations combined with KMC simulations \cite{Schuwalow14}. The values for the diffusion activation energies extracted from the Arrhenius plot can directly be compared with experimental measurements of tracer diffusion coefficients. Such simulations also allow directly comparing the mobility of alloying elements in Ni and Co \cite{Neumeier-08} as a quantity of interest in the
investigation of Ni- and Co-base superalloys.
Furthermore, the simulations provide insight into the mobility of vacancies and how the presence of alloying elements influences vacancy transport \cite{Schuwalow14,Goswami14,Grabowski18,Goswami19}. In complex alloys, the diffusion properties depend on the composition. An accurate description on an atomistic level requires taking into account interaction energies between solute atoms, as well as their influence on microscopic diffusion barriers \cite{Grabowski18,Goswami19}. This also applies to chemically ordered alloys such as the Ni$_3$Al $\gamma'$\;phase where diffusion processes might take place on the same sublattice or cause a swap between sublattices \cite{Duan07,Gopal12,Zhang14b}.
In multicomponent alloys, an accurate mapping of all possible diffusion processes becomes rather involved.
Here, KMC simulations can be combined with more advanced approaches such as a cluster expansion of the diffusion barriers parameterized by electronic structure calculations \cite{VanderVen01}.

\section{Line defects}

Line defects, here mainly dislocations, can be studied both by DFT calculations and atomistic simulations with TB/BOP models or classical potentials.
The small number of atoms that can be treated with DFT approaches, however, requires advanced methods for the boundary conditions and allows only for static calculations \cite{Woodward2008,Tan19}.
The motion of extended dislocations, as well as their interactions with each other and other defects can, so far, only be simulated using interatomic potentials with their known limitations in accuracy and chemical complexity.

\subsection{Dislocation core structure}

\begin{figure}[htb]
\includegraphics[width=0.42\columnwidth]{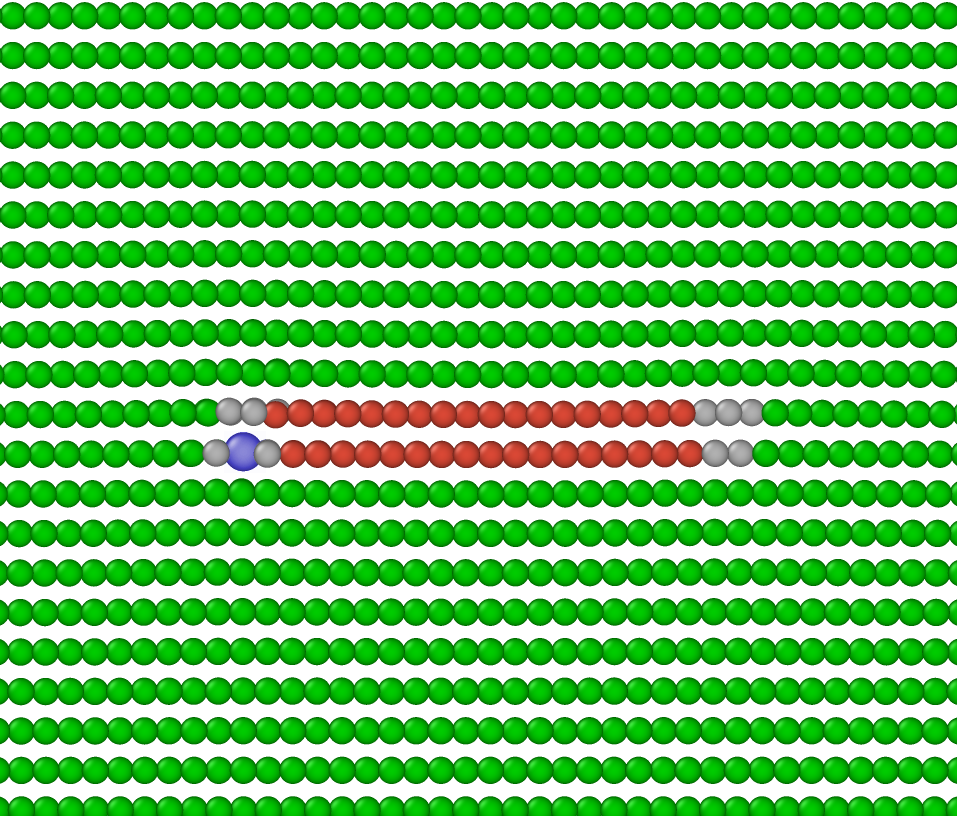}
\includegraphics[width=0.48\columnwidth]{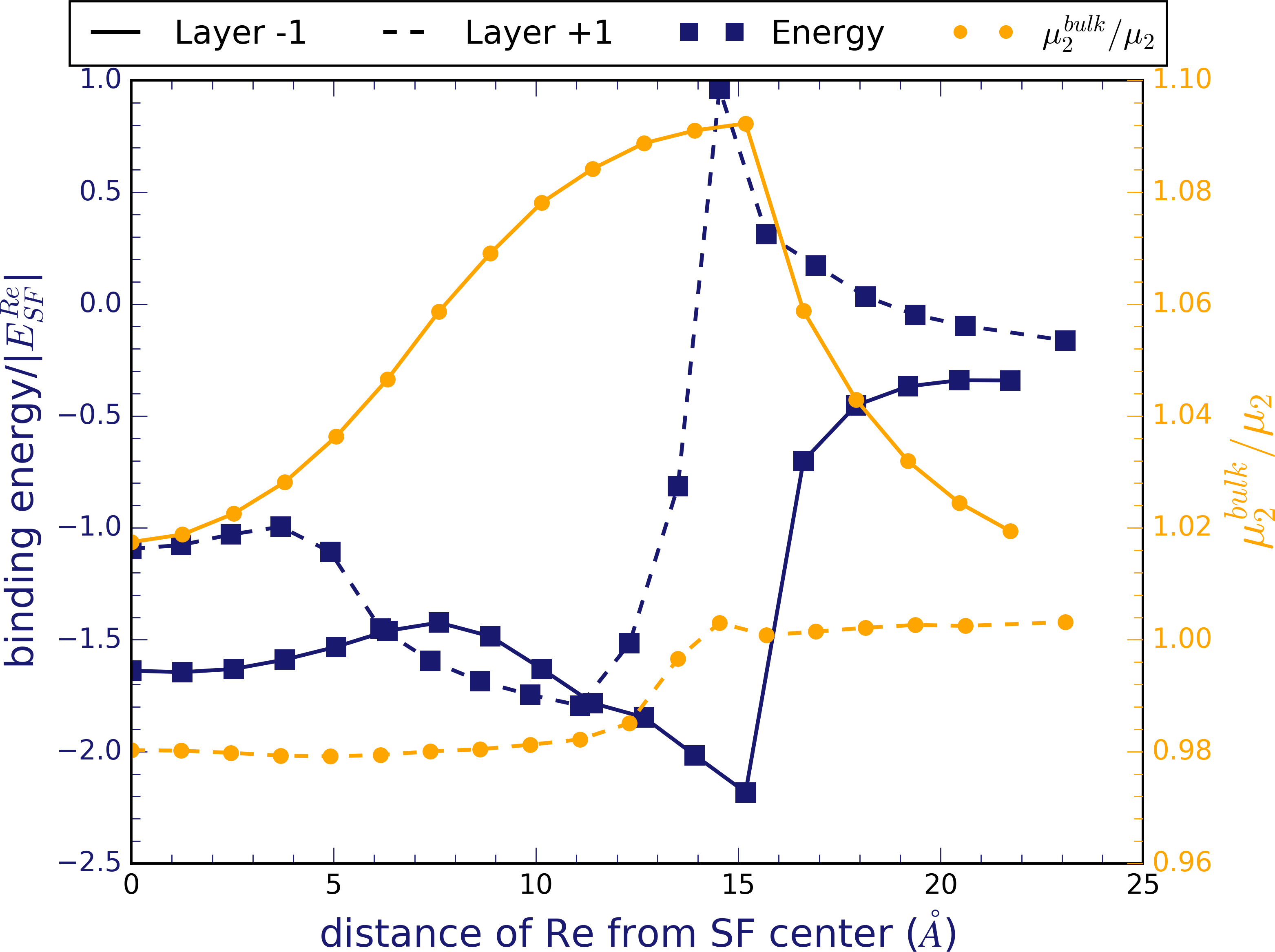}
\caption{\small Atomistic simulation of a Re atom (blue) in an edge dislocation complex of two partial dislocations (grey) and a stacking fault (red) in fcc Ni (green). This structure is obtained by removing a half-plane of atoms from fcc Ni, subsequent relaxation of the atomic positions which leads to the partial dislocations, replacement of one Ni atom by Re, and further atomic relaxation. (Right) The interaction energy (blue) of the single Re atom at different positions relative to the dislocation complex shows that the energetically most favorable position is in the tensile layer (layer -1) of the partial dislocation. In this position, the Re atom has the maximum accessible volume factor (orange) as obtained by TB and BOP calculations. Figure adapted from \cite{Katnagallu-19}.}
\label{fig:NiRe-SF}
\end{figure}

Dislocations in pure Ni as model systems for the $\gamma $\;phase have been extensively studied
\cite{Szelestey2003,Bitzek2005,Bianchini2016,Tan19} also with respect to segregation of solute atoms \cite{liu-wang-17,Katnagallu-19}. In atomistic simulations, these dislocations are generated by displacing the atoms according to an analytic solution of the displacement field around the dislocation, followed by a relaxation of the atomic positions. For an edge dislocation this leads to two partial dislocations separated by an intrinsic stacking fault as shown in Fig.~\ref{fig:NiRe-SF}.
Dislocations in the $\gamma'$\;phase are much more diverse, including partial dislocations bounding different planar defects (see Section~\ref{chap13sec:interfaces}), different possible glide planes and Burgers vectors \cite{PhysicalMetallurgy}.
Consequently, most atomistic studies were directed to reproduce experimentally observed dislocation structures and to determine their properties.

Understanding $\langle $110$\rangle $ screw superdislocations in Ni$_3$Al is of particular importance due to their potential for forming Kear--Wilsdorf (KW) lock configurations \cite{KW} and thus leading to the anomalous yield behavior of L1$_2$ alloys.
Therefore, much effort has been invested during the 1980s and 1990s to investigate their core structure \cite{Yamaguchi1982,Paidar1982,Yoo1986,Yoo1987,Farkas1988,Yoo1989,PPV,Par96,Wen1997,Wen1998}.
Depending on the simulation details, different core structures are possible, including dissociation of the superdislocation on a $\{111\}$ plane into a pair of $1/2 \langle 101 \rangle$ superpartial dislocations separated by an antiphase boundary (APB) and nonplanar configurations.
The detailed locking mechanism is commonly believed to involve the thermally-activated formation of Paidar--Pope--Vitek (PPV) locks \cite{PPV}, which has been recently studied in detail \cite{Par96,Ngan2004a,Rao2012}, including the use of NEB calculations to calculate activation energies \cite{Ngan2004a,Rao2012}.
In this context, it is important to note that the activation barriers depend strongly on the studied configuration. Considering, e.g., the interaction with forest dislocations can significantly reduce the activation energy for the critical cross-slip process \cite{Rao2012}.
Recent atomistic simulations suggest furthermore, that under applied stress and elevated temperatures also $\langle 110 \rangle $ edge superdislocations can show nonplanar dissociation leading to complex Lomer--Cottrell locks \cite{Xie2011,Xie2013,Fan2016}.

In addition to the usual $\langle $110$\rangle $ superdislocations also $\langle 100\rangle $ edge superdislocations were observed to penetrate the $\gamma'$\;particles, in particular during high temperature, low stress creep,
where this process is believed to be the rate limiting step \cite{Eggeler1997,Srinivasan2000_100,Link2005}.
The dissociation of a perfect $a \langle 100\rangle \{010\}$ edge superdislocation in Ni$_3$Al was studied by Kohler et al. \cite{Kohler2006},
who found that it could dissociate in a symmetric dissociation similar to the Hirth lock in fcc structures, an asymmetric dissociation and a dissociation into two interlocked $a/2\langle 110\rangle $ dislocations, with the
Hirth lock being the most stable configuration.
HRTEM observations and image simulations by Srinivasan et al. \cite{Srinivasan2000_100}, however, showed a somewhat different core structure, which might be related to the coupled glide-climb process necessary for the motion of $\langle 100\rangle $ dislocations.

\subsection{Dislocation mobility and solid solution strengthening}

The Peierls stress $\tau _{P}$ necessary to initiate dislocation motion can be easily determined by quasistatic calculations on infinite straight dislocation lines in a slab geometry with periodic boundary conditions subjected to different shear stresses.
The same setup can be used to measure the temperature-dependent drag coefficient $B$ caused by the interaction of phonons with the moving dislocation by performing MD simulations at different temperatures.
For pure Ni, these parameters have been determined by Bitzek et al. \cite{Bitzek2004b,Bitzek2005}.
Alternatively, the Peierls stress can also be determined by DFT calculations of the generalized stacking-fault energy in a Peierls--Nabarro model as, e.g., in \cite{Wang2014} for Ni$_3$Al.
Although an important parameter in discrete dislocation dynamics (DDD) simulations, the drag coefficient $B$, see Chapter~12,
for dislocations or superdislocations in Ni$_3$Al has not yet been determined by atomistic simulations.

The strengthening effect of different solutes can be predicted, e.g., by combining DFT calculations of solute--dislocation interaction energies with a Labusch-type model \cite{Leyson2010}.
Such an approach has, however, not yet been used to calculate solid solution strengthening in
the $\gamma $\;or $\gamma'$\;phases of Ni-base superalloys.
Alternatively, direct MD simulations of dislocation motion in solid solutions can be used to determine parameters like static and dynamic threshold stresses and effective ``friction coefficients'' for dislocations, including superdislocations in $\gamma $\ \cite{Rodary2004,Marian2006,Zhang2013g,Fan2016}.
This approach depends critically on the availability and quality of atomic interaction potentials.
Although recently simulations have focused on concentrated solid solutions \cite{Rao2017},
no simulations were so far performed on realistic model systems for Ni-base superalloys.
Please also refer to Chapter~14 regarding the modeling of solid solution strengthening in crystal plasticity.

It needs to be stressed that the above approaches to study dislocation mobility neglect diffusive processes.
The shearing of the $\gamma'$\;phase by partial dislocations, however, requires reordering steps that involve short-range diffusion \cite{Kovarik-09,Zhou2011}.
Modeling such concerted diffusive--displacive processes at the atomic scale remains one of the fundamental challenges for atomistic simulations \cite{Li2011_DMD}.
The study of dislocation climb by atomistic simulations is therefore in its infancy \cite{Sarkar2012_DMD,Baker2016}, and the effects of solute atoms on the climb
mobility are yet to be investigated at the atomic scale.

\section{Interfaces and planar defects}
\label{chap13sec:interfaces}

The calculation of the $\gamma$/$\gamma'$ interface of a superalloy with realistic chemical complexity requires large supercells and proper combinatorial sampling of the disordered $\gamma $\;phase.
A common approximation is therefore to mimic the $\gamma $\;phase by pure Ni which has been used successfully in calculations with classical EAM (embedded-atom method) potentials of different interfaces \cite{Mishin2014} and in DFT calculations of the influence of alloying elements on the interface energy \cite{liu-AAAA-liu-17}.

The mechanisms of the plastic deformation of superalloys in the different pressure and temperature regimes are discussed in detail in
Chapter 6.
The plastic deformation is governed by $\gamma'$\;cutting processes (see, e.g., \cite{Agudo-14}) and by microtwinning (see, e.g., \cite{Kovarik-09,Barba-17}).
These involve several steps including dislocation dissociation into partials and the movement of the partials, as well as the segregation to the partial and to stacking faults between partials and behind the trailing partial.
The most important planar faults in the context of shearing deformation are superlattice intrinsic stacking faults (SISF), superlattice extrinsic stacking faults (SESF), antiphase boundaries (APB), and pairs of twin boundaries.
These phenomena at the atomic level are ideal candidates for atomistic simulations, particularly regarding clarification of
(i) the relative formation energy of different planar faults (see, e.g., \cite{Titus-15-1}),
(ii) the influence of alloying elements on the fault energy, and
(iii) the segregation of alloying elements to the fault plane (see, e.g., \cite{Viswanathan-15,Eggeler-16}).

The majority of recent calculations for planar faults are, in fact, performed on the basis of DFT calculations due to a lack of reliable alternatives for multicomponent systems.
The stacking faults are accessible to atomistic simulations by supercell calculations and, in few cases, by an approximation with an axial Ising model (AIM) \cite{Denteneer-87}.
Disorder of the atoms on the crystal lattice (due to, e.g., off-stoichiometry or finite temperature) can be introduced by special quasirandom structures (SQS) \cite{Zunger-90}, by cluster expansions (CE), or by mean-field approaches like the coherent phase approximation (CPA).

The energy of the APB that is formed during $\gamma'$\;cutting of a dislocation is of central importance for shear resistance, see Chapter~6.
Several recent works, including supercell DFT calculations \cite{Chandran-11,Vamsi-14,Crudden-14,Kumar-18-2}, DFT-based CE \cite{Sun-17}, and DFT-based CPA \cite{Gorbatov-16},
showed consistently that alloying with group IV (e.g., Ti) and group V (e.g., Ta) transition-metal elements can considerably increase the APB energy and hence the shear resistance.
Corresponding works for SISF and SESF were carried out with supercell DFT calculations \cite{Wen-12,Kumar-18-2} and AIM-based approximations \cite{Breidi-18}.
Considerably fewer works are available that performed corresponding calculations for the microtwinning mechanism during creep.
The segregation of selected alloying elements to SESF and SISF was determined by supercell DFT calculations \cite{Eurich-15} that were recently extended and interpreted in terms of reordering kinetics and intermediate fault structures \cite{Rao-18}.

\section{Microstructure and defect--defect interactions}

Simulating the $\gamma$/$\gamma'$ microstructure of Ni-base superalloys, its deformation, and final fracture, as well as the direct interaction between defects, requires atomistic simulations with large numbers of atoms that presently can only be realized with semiempirical potentials.
Such simulations are very useful in further developing our mechanistic understanding of deformation processes and can help with the interpretation of experimental observations, which due to their limited time resolution might not capture all relevant details.

\subsection{Deformation of individual precipitates}

Although not directly related to the study of superalloys under typical application conditions,
recent experimental \cite{Landefeld2012,Maass2012} and MD simulation \cite{Amodeo2014,Shreiber2015,Houlle2018a}
studies on individual $\gamma'$\;nanocubes under compression allowed to investigate the deformation behavior of the pure $\gamma'$\;phase.
These defect-free cubes deform by the nucleation of Shockley partial dislocations, leaving behind complex stacking-faults that can at larger strains transform into a pseudotwin structure \cite{Amodeo2014}.
The detailed deformation mechanisms, however, depend critically on the used potentials \cite{Shreiber2015},
showing the importance of performing well-controlled experiments to
validate interatomic potentials.
Combining such well-defined experimental or simulation studies on the individual phases,
as well as on the $\gamma$/$\gamma'$ microstructure can help elucidate the influence of
constraints and misfit stresses on the mechanical performance of single-crystalline superalloys \cite{Houlle2018a}.

\subsection{Interfacial dislocation networks}

The misfit dislocation network (MFDN) that forms directly upon energy minimization due to the lattice misfit between the $\gamma $\;and the $\gamma'$\;phases has been the subject of many atomistic simulations \cite{Zhu2005,Yashiro2007,Xie2009,Wu2011c,zhu2013,Yu2015,Ye2015,APT,Li2018a,Ding2018}.
These studies have been performed on systems in which the $\gamma $\;phase is represented by pure Ni, whereas the $\gamma'$\;phase is modeled by stoichiometric Ni$_3$Al.
The resulting lattice misfit of $\delta \approx 1.44$--$2.8\%$, depending on the potential, is one order of magnitude larger than in typical superalloys, which furthermore have a negative misfit \cite{Parsa2015}.
Only recent studies included alloying elements \cite{Ding2018} and nonstoichiometric compositions \cite{APT}.

\begin{figure}[h]
\includegraphics[width=0.9\columnwidth]{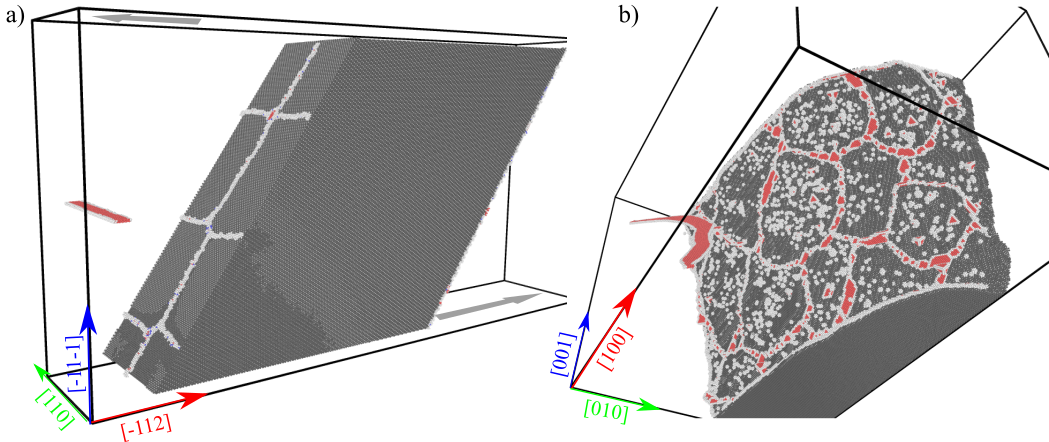}
\caption{Simulation setups used to study dislocation--precipitate interactions: (a) typically used setup to determine critical stresses for precipitate cutting, e.g., in \cite{zhu2013}, where stresses and/or displacements can be applied on the top and bottom surfaces (symbolized by the grey arrows). These setups with perfectly planar interfaces lead, however, to a square MFDN with unrealistic compact cores; (b) experimentally-informed simulation setup using the precipitate morphology obtained from APT data \cite{APT} with extended MFD cores as observed in experiments (only atoms belonging to defects or the precipitate are shown, atoms in red are part of a stacking fault).}
\label{fig5:dislo-precipitate}
\end{figure}

The most important critique of these simulations is, however, their artificial construction of
the interfacial dislocation network, often using perfectly planar interfaces.
In reality, the interfacial dislocation network (IDN) forms by deposition of channel dislocations under creep conditions and their subsequent rearrangement \cite{Link2005}.
In this respect, the artificial MFDN can only be seen as an idealized arrangement of interfacial dislocations that would most effectively reduce the misfit stresses.
However, in most simulation studies, the IDN dislocations show a compact core, see Fig.~\ref{fig5:dislo-precipitate}(a).
This is due to the fact that the $\{100\}$ interface plane is not a natural glide plane for dislocations in the fcc crystal structure.
By using realistic interface morphologies, it was recently shown that dislocations in the MFDN
assume configurations in which they spread out on $\{111\}$ planes, and can form stair-rod junctions as also observed in multiple experiments \cite{APT}.
In addition, also the Burgers vectors in the MFDN changed due to the interface curvature \citep{APT}.
An example for a realistic MFDN can be seen in Fig.~\ref{fig5:dislo-precipitate}(b).

\subsection{Dislocation--precipitate interactions}

The interaction of $\gamma $\;matrix dislocations with $\gamma'$\;precipitates is the main reason for the high creep strength of Ni-base superalloys.
Here, atomistic simulations can in principle be used to identify the conditions under which single dislocations or superdislocations can cut into the $\gamma'$\;phase.
The interaction between single, infinitely long edge, or screw dislocations with a regular array of spherical $\gamma'$\;precipitates with diameters up to 6~nm was studied by Proville and Bako \cite{Proville2010a}.
They showed a transition from dislocation cutting to Orowan bowing with increasing precipitate diameter.
Kohler et al. \cite{Kohler2005} used a similar methodology but focused on even smaller precipitates.
The more relevant case of studying the interaction of super dislocations with spherical precipitates was recently simulated by Hocker et al. \cite{Hocker2019} and Kirchmayer et al. \cite{AEM2020}.
The former group used a similar setup as in \cite{Proville2010a,Kohler2005} using regular arrays of spherical precipitates. In that case it was found that, depending on whether the distance between the superpartial dislocations is larger or smaller than the precipitate diameter (corresponding to weak or strong coupling between the
superpartial dislocations \cite{Huther1978}), different partial dislocations govern the critical resolved shear stress (CRSS) to pass the precipitate.
Kirchmayer et al., however, used realistic precipitate morphologies and arrangement obtained from atom probe tomography (APT), and were thus able to show that for a relatively wide distribution of precipitate sizes
weak and strong pair coupling can be at play simultaneously~\cite{AEM2020}.
Spherical precipitates were also studied by Takahashi et al. \cite{Takahashi2011} and Kondo et al. \cite{Kondo2018}, however, in their simulations the precipitates consisted of the $\gamma $\;phase which were cut by
superdislocations from the surrounding Ni$_3$Al $\gamma'$ phase.
Besides providing information about the dislocation--precipitate interaction processes, such simulations can provide -- with the usual limitations of the unrealistic compositions -- quantitative information on
precipitate cutting stresses $\tau _{c}$ and on the relative importance of, e.g., the coherency stresses, the APB energy or the energy to create an interface step.

Single-crystalline superalloys are strengthened by the presence of large, cuboidal $\gamma'$ precipitates that force the dislocations to glide in narrow channels.
Relatively few detailed atomistic studies of dislocation--precipitate interactions in this type of microstructure have been reported so far.
Such studies would, however, be important to inform discrete dislocation dynamics simulations, see \cite{Yashiro2006b} and Chapter~12,
in particular regarding, e.g., the influence of local interface curvature and chemical composition gradients on dislocation--precipitate interactions.
The motion of dislocations into $\gamma $\;channels bounded by two cuboidal $\gamma'$\;precipitates was first studied by Yashiro et al. \cite{Yashiro2002}.
Recently, Xiong et al. \cite{Xiong2017} performed a quantitative study to determine the stress required for matrix dislocations to penetrate into the channel in a similar setup, however, with a preexisting MFDN.
Overall, they found an inverse proportionality between the critical stress and the channel width.
The influence of the MFDN on the interaction of channel dislocations with the $\gamma'$\;precipitate was recently studied using different setups \cite{zhu2013,APT,Xiong2017}, see also Fig.~\ref{fig5:dislo-precipitate}.
The simulations with a realistic MFDN and dislocation core structures resulting from an APT-informed sample showed that in particular the colinear interaction of the first channel dislocation with a misfit dislocation protects the precipitate from being sheared by a superdislocation as due to the dislocation annihilation no trailing superpartial is available \cite{APT}.
Prakash and Bitzek \cite{Prakash2017} furthermore studied the interaction of dislocation loops with various shapes and arrangements of $\gamma'$\;precipitates.
The different stress fields caused by the misfit stresses lead to significant differences regarding the cutting of precipitates and the penetration of the dislocation into the channels, demonstrating the importance of taking deviations from the idealized morphologies and topologies into account.

\subsection{Deformation and fracture of superalloys}

\begin{figure}[h]
\includegraphics[width=0.9\columnwidth]{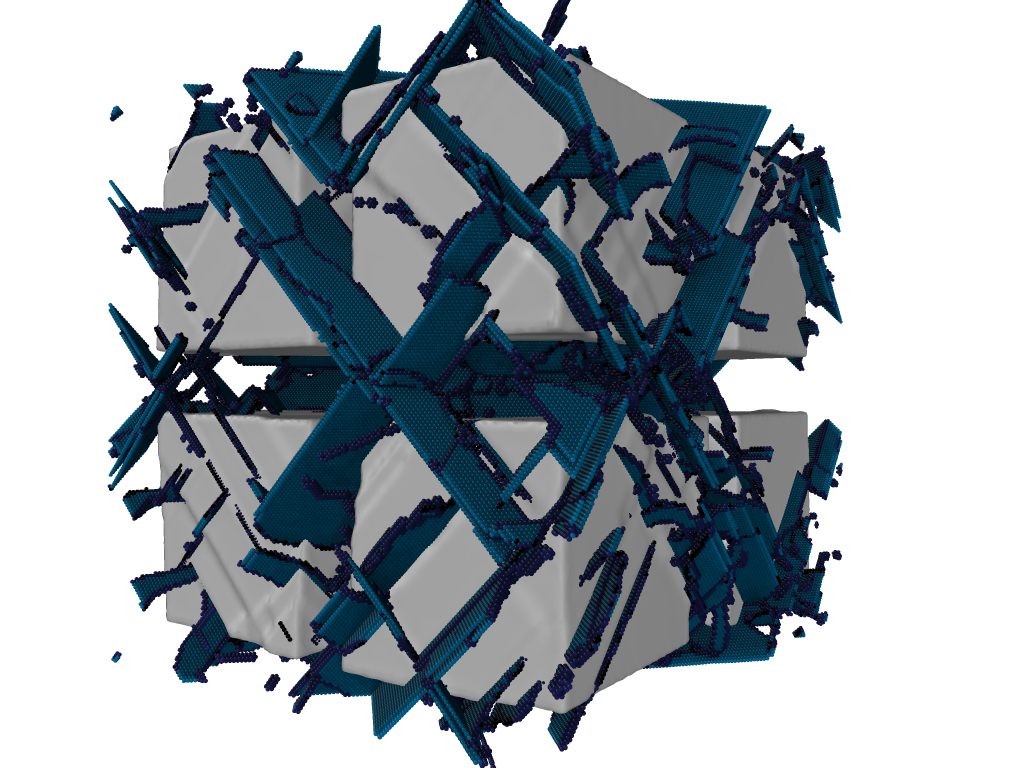}
\caption{Dislocation structure after 8.4 \protect \% uniaxial tensile strain at $T=1250$ K with the EAM potential by Mishin \cite{Mishin-04} (the structure is periodic in all directions, cube side length 16.4 nm, channel width 4.6 nm, 6.9 million atoms). 
Only atoms belonging to the dislocation cores and stacking faults as well as the shape of the $\gamma'$ cubes are shown.}
\label{fig6:tensiletest}
\end{figure}

Interestingly, relatively few large-scale MD studies exist on the plastic deformation and stress--strain response of superalloys with $\gamma$/$\gamma'$ micro\-structures. Ma et al. \cite{Ma2016b} performed tensile tests on an Ni/Ni$_3$Al microstructure, however, made up of an unrealistic checker-board arrangement of cubic $\gamma $\;and $\gamma'$\;crystallites. Using one cubic $\gamma'$ precipitate surrounded by $\gamma $, Li et al. \cite{Li2018a} showed a tension compression antisymmetry in the stress-strain response. The result of a recent MD simulation with eight $\gamma'$\;cubes arranged according to an idealized $\gamma$/$\gamma'$ microstructure is shown in Fig.~\ref{fig6:tensiletest}.
All these simulation results are, however, of limited relevance as under these conditions, the deformation is governed by the homogeneous nucleation of dislocations at extremely high strain rates of the order of $10^{7}$ to $10^{9}$ per second.
This is in stark contrast to typical experimental conditions and samples that contain preexisting ingrown dislocations.
This highlights the need for a multiscale approach in which the information from atomistic simulations are used in, e.g., DDD simulations at lower strain rates, larger sizes, and with more realistic initial dislocation densities. In particular, for the later stages of deformation where creep or fatigue fracture becomes relevant, atomistic simulations are uniquely positioned to provide criteria for crack advance or dislocation nucleation to meso- and continuum scale models \cite{Bitzek2015,Moller2018}.
The corresponding atomistic studies regarding crack nucleation and crack propagation in $\gamma $/$\gamma'$ microstructures \cite{Liu2013,Yu2014,Shang2018} are, however, still in their infancy.

\section{Limitations of atomistic simulations for superalloy design}

While progress in atomistic theory in the past years means that many properties of superalloys became accessible, other aspects of superalloy performance are not directly accessible by atomistic simulations. Here we summarize the main limitations of present day atomistic simulations for superalloys.
\begin{itemize}
\item While DFT is used heavily in atomistic simulations for superalloys, the approximations in current exchange correlation functionals still limit the use of DFT data for alloy design. For example, formation energies and enthalpies computed by DFT can in principle be included in CALPHAD assessments, but accuracy of the DFT data, compatibility to experimental data sets and exhaustive DFT data for multicomponent alloys remain issues that need to be resolved together with an update of the CALPHAD approach to bring it closer to atomistic simulations. The same holds true for elastic constants and interface energies that are required for phase field simulations.
\item Superalloys are multicomponent materials. Interatomic potentials for multicomponent materials are not available today. This means that atomistic simulations often cannot provide input for mesoscale or continuum models. The development of quantitatively accurate multielement interatomic potentials remains one of the grand challenges in the field.
\item The time scale of atomistic molecular dynamics simulations is limited by the time step for the numerical integration of the atomic trajectories. The time step is on the order of $10^{-15}$\,s. This means that time scales required for studying processes that include diffusion, e.g., dislocation climb-glide creep, are often not accessible. Expanding atomistic simulations to diffusive time scales is one of the major challenges for the field.
\item An even greater gap exists between atomistic simulations of microstructural events and the modeling of plasticity. Models of plasticity need to define their parameters explicitly in such a way that atomic-scale simulation procedures may be developed to provide the required parameters.
\item Continuum models are sometimes incompatible or at least not directly compatible with atomistic simulations. The transfer of results from atomistic simulations to continuum descriptions remains one of the roadblocks for a first principles guided design of superalloys.
Both the atomistic and continuum modeling communities need to make great strides towards a more efficient exchange.
\end{itemize}

\section{Summary: modeling Ni-base superalloys from electrons to microstructures}

The rapid progress in atomistic modeling and simulation over the past years brings first-principles computational design of superalloys into reach. Today, atomistic modeling enables the prediction of key aspects of superalloy properties, from solute bond chemistry to microstructural properties. We reviewed the different representations of the interatomic interaction that are commonly used for modeling superalloys, briefly introduced atomistic simulation methods in the context of superalloys, and outlined properties of superalloys that can be computed by atomistic simulations.

Limitations of atomistic simulation methods means that many properties of superalloys are not accessible today and the development of ``approximate practical methods'' that Paul Dirac envisioned nearly a century ago will remain a focus of the atomistic simulations community.

We hope that the present chapter may induce discussions and collaborations that will contribute to further narrow the gap between atomistic simulation and superalloys, and in this way help advance the design of novel superalloys.

\section*{Acknowledgement}
We are grateful to A.P.A. Subramanyam, H. Lyu, F. Houll\'{e}, A. Prakash, and J.J. M\"{o}ller for their help in preparing some of the figures. 
We acknowledge financial support from the German Research Foundation (DFG) through projects C1, C2, and C3 of the collaborative research center SFB/TR 103.

\bibliography{chapter11}

\end{document}